\documentclass[twocolumn,showpacs]{revtex4}

\usepackage{graphicx}%
\usepackage{dcolumn}
\usepackage{amsmath}

\makeatletter
\def\btt#1{\texttt{\@backslashchar#1}}%
\DeclareRobustCommand\bblash{\btt{\@backslashchar}}%
\makeatother

\begin{document}

\preprint{HEP/123-qed}

\title[Short title]{A switchable controlled-NOT gate in a spin-chain NMR quantum computer
}% Force line breaks with \\

%\author{A. Goto et al.}
\author{Atsushi Goto} 
 \email{GOTO.Atsushi@nims.go.jp}
% \thanks{}%Lines break automatically or can be forced with \\
\author{Tadashi Shimizu}%
% \email{SHIMIZU.Tadashi@nims.go.jp}
\author{Kenjiro Hashi}%
% \email{HASHI.Kenjiro@nims.go.jp}
\author{Hideaki Kitazawa}%
% \email{KITAZAWA.Hideaki@nims.go.jp}
\affiliation{Nanomaterials Laboratory, National Institute for Materials Science, Sakura, Tsukuba, 305-0003, Japan}
%\affiliation{CREST, Japan Science and Technology Corporation, Kawaguchi, Saitama, 332-0012, Japan}
%

\date{\today}% It is always \today, today, but you may specify any date with \date.

\begin{abstract}
A method of switching a controlled-NOT gate in a solid-stae NMR quantum computer is presented.
Qubits of $I=1/2$ nuclear spins are placed periodically along a quantum spin chain (1-D antiferromagnet) 
having a singlet ground state with a finite spin gap 
to the lowest excited state caused by some quantum effect.
Irradiation of a microwave tuned to the spin gap energy 
excites a packet of triplet magnons at a specific part of the chain
where control and target qubits are involved.
The packet switches on the Suhl-Nakamura interaction between the qubits,
which serves as a controlled NOT gate.
The qubit initialization is achieved by a {\it qubit initializer} 
consisting of semiconducting sheets attached to the spin chain,
where spin polarizations created by the optical pumping method in the semiconductors 
are transferred to the spin chain.
The scheme allows us to separate the initialization process from the computation,
so that one can optimize the computation part without being restricted by the initialization scheme,
which provides us with a wide selection of materials for a quantum computer.
\end{abstract}


\maketitle

%\tableofcontents

\section{Introduction}
\label{sec:introduction}

A quantum computer (QC) is a Turing machine 
which performs information processing 
based on the principles of quantum mechanics. 
It makes the best use of the features of the quantum mechanics, 
such as superposition and entanglement of quantum states.
These features enable us to perform parallel computation 
for all the possible states simultaneously,
which makes it possible to deal with the problems 
that are formidable for classical (conventional) computers.
New algorithms recently discovered 
\cite{deutsch85,shor97,grover98} 
have shown the great promise of the QC's,
which has accelerated the attempts to implement the QC's in actual physical systems.

A QC is composed of a set of two-level systems called qubits.
The qubits need to be isolated enough from the environment, and controllable
from outside to deal with the information. 
In this respect, nuclear spin systems in a matter are the promising candidates, 
because they are only weekly coupled with the environment 
(electron systems) through hyperfine couplings,
and controllable by the well-established technique of the nuclear magnetic resonance (NMR).
Actually, the first 2-qubit QC's were implemented by solution NMR
\cite{chuang98, jones98}.
The quite successful implementations proved the great promise of the NMR-QC's.
It is unfortunate, however, that the solution NMR-QC has a difficulty in its scalability.
The number of available qubits in solutions is limited 
because of the limited number of nuclei in one molecule.
The required number of qubits with which QC can surpass its classical analogue 
is estimated to be more than $10^3$,
so that it is a primary concern to increase the number of qubits.

So far, two models have been proposed to systematically increase the number of qubits 
in the NMR-QC's.
Kane proposed a multi-qubit NMR QC model, 
which utilizes $^{31}$P nuclei embedded in a Si matrix
\cite{kane98}.
This model provides us with a chance to extend the number of qubits systematically, 
whereas it requires fine structures to be fabricated and controlled, 
realization of which is a challenge to the current state-of-the-art nano-technology.
On the other hand, Yamaguchi and Yamamoto proposed to utilize nuclei 
as they are in nature, i.e., in crystals 
\cite{yamaguchi99}.
The $^{31}$P nuclei in CeP are placed in the magnetic field gradient, 
so that each nucleus (qubit) can be accessed by adjusting the NMR frequency.
The proposal is very attractive because of its simple structure, 
although it still has some technical obstacles to be cleared 
\cite{hashi00}.
We have pursued this possibility checking the key issues.

\begin{figure}
\includegraphics[scale=0.55]{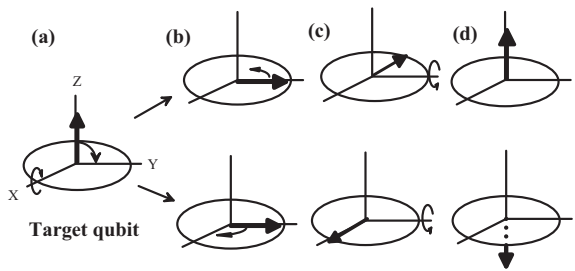}% Here is how to import EPS art
\caption{
Schematic view of the sequence for the c-NOT gate in the rotating frame of the target qubit. 
Upper and lower figures correspond to the cases of the up (0) and the down (1)
states of the control qubit. 
(a) The target qubit spin is pointing to the Z-direction.
The spin is rotated by a $\pi/2$ pulse around the negative X axis.
(b) The spin starts to turn in the XY plane due to the additional field caused by the control qubit. 
(c) The spin turns by $\pm 90^o$ to positive or negative X direction in the XY plane. 
(d) The spin is rotated again by a $\pi/2$ pulse around the positive Y axis.  
The direction of the spin with respect to Z is hereby
controlled according to the spin state of the control qubit.
}
\label{CNOT-gate}
\end{figure}

One of the key issues is how to provide an inter-qubit (inter-nuclear) coupling,
which is used for a controlled NOT (c-NOT) gate shown in Fig. \ref{CNOT-gate}.
So far, a nuclear dipole (direct) coupling has been supposed 
to be a potential inter-qubit coupling
\cite{yamaguchi99, ladd00}. 
It is unfortunate, however, 
that the dipolar coupling is always present whenever qubits are put close to each other, 
so that one should continue applying the decoupling sequences to remove unwanted couplings.
Apparently, these sequences consume a great deal of time, 
causing a longer computation time. 
Moreover, as the number of qubits is increased,
the number of inter-nuclear couplings is increased accordingly,
so that the decouplings become more and more complicated,
and they eventually become formidable.
So, a decoupling-free QC is highly desired, i.e., 
the inter-nuclear coupling should be able to be switched on when, 
and only when necessary.

Another key issue is the method of distinguishing, in the frequency domain, 
each nucleus (qubit) separated by a distance of the order of a lattice constant. 
Even the field gradient of 1T/$\mu$m
\cite{goldman00}
is still marginal for this purpose
\cite{ladd00}.
The larger inter-nuclear (qubit) distance in the real space makes 
the distance between the adjacent NMR lines wider in the frequency domain, 
which relieves this constraint.
Hence, the inter-nuclear coupling should be able to reach rather long distance. 
Unfortunately, the direct nuclear dipole coupling reaches at most a few lattice points
\cite{goto02a}.

These facts motivate us to seek for the possibility 
of the long range indirect couplings mediated by electrons. 
The indirect couplings include the J-coupling due to the covalent bondings, 
the RKKY interaction in metals
\cite{ruderman54}, 
and the Suhl-Nakamura (SN) interaction in magnets
\cite{suhl58, nakamura58}.
Among them, the SN interaction has the characteristics preferable for the present purpose,
such as the long-range nature of the coupling and 
the external controllability of the coupling strength.

In this paper, we present details of the model of a solid-state NMR-QC 
with an inter-qubit coupling provided by the SN interaction
\cite{goto02b}.
A singlet-triplet transition in a quantum spin chain (1D antiferromagnet)
provides us with a switch for the inter-qubit coupling.
The model has the following advantages over the other existing models.
(1) It is intuitive because the computation starts in the silent environments 
rather than the turbulence of interactions,
which makes the designs of the logic gates simpler.
(2) The long-range nature of the SN interaction 
allows us to place qubits apart farther from one another,
which facilitates distinguishment of the qubits in the frequency domain.
(3) The presence of the spin gap in the quantum spin chains 
allows simple operation of the gate switching 
compared to that proposed for the Kane's type NMR-QC
\cite{khitun01}.
We also present the scheme of {\it qubit initializer}, 
an effective nuclear polarizer 
comprising the optical pumping and the polarization transfer methods. 
The scheme enables us to separate the materials 
responsible for the initialization and computation, 
so that one can optimize the computation part
without being restricted by the initialization scheme.

\section{Model}
\label{sec:model}
\begin{figure}[tb]
\includegraphics[scale=0.55]{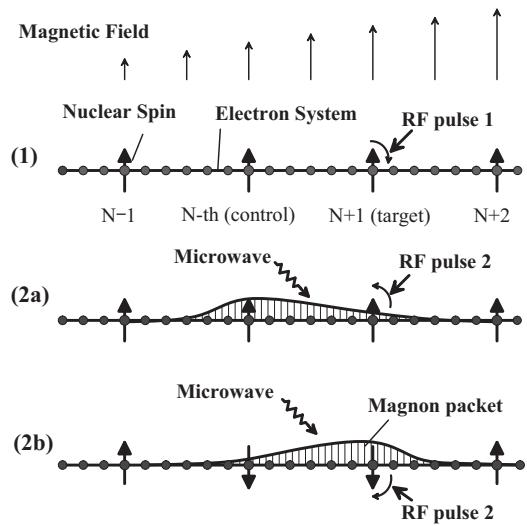}% Here is how to import EPS art
\caption{
Schematic illustration of the c-NOT-gate switching
in a 1D antiferromagnet in a magnetic field gradient.
The $I=0$ and $1/2$ (= qubits) nuclei are shown by balls and arrows, respectively.
(1) All the inter-nuclear couplings are switched off in the absence of magnons.
A $\pi/2$ pulse is applied to the target ($I_{N+1}$) qubit.
(2) A magnon packet (hatched part) is excited 
between the control ($I_N$) and the target ($I_{N+1}$) qubits.
The magnon packet entangles the two qubits, and the second $\pi/2$ pulse 
makes the $I_{N+1}$ spin turn either back to the +1/2 state for $I_{N}=1/2$ (2a), 
or forth to the -1/2 state for $I_{N}=-1/2$ (2b) (see Fig. \ref{CNOT-gate}).
The (2a) and (2b) states are superposed in the actual computation.
}
\label{j-switch}
\end{figure}
Here, we present the outline of the model.
The main idea is illustrated in Fig. ~\ref{j-switch}.
The system consists of a one-dimensional array of electron spins (quantum spin chain) 
placed in a magnetic field gradient, 
which is produced by a micromagnet fabricated outside the spin chain
\cite{goldman00}.
Suppose that the electron spins are paired into singlets in the ground state 
($\mid s s_z\rangle$=$\mid0 0\rangle$) with a finite gap 
to the lowest triplet branch ($\mid1 -1\rangle$) of the $k=0$ magnon modes
because of some quantum effects.
Examples of such situations can be found in spin ladder, Haldane,
dimer and spin-Peierls systems.
Also suppose that nuclei ($I=1/2$) serving as qubits can be placed periodically, 
e.g., every ten lattice points, 
each of which has a hyperfine coupling with the electron spins.

Since there are no unpaired electron spins in the ground state, 
the nuclear spins are well-isolated from the environment. 
The nuclei are also decoupled from the charge and lattice properties of the electrons 
because $I=1/2$. 
The rather long distance between qubit nuclei is effective 
both to diminish the direct nuclear dipole couplings between qubits
and to distinguish one qubit from another in the NMR frequency domain,
because the longer inter-qubit distance leads to 
larger interval between NMR lines in the given field gradient.

It is known that virtual exchanges of magnons between nuclear spins 
result in the inter-qubit interaction called the Suhl-Nakamura (SN) interaction
\cite{suhl58, nakamura58}. 
Since it is not necessarily accompanied by actual excitations of magnons, 
it can exist even at temperatures well beneath the excitation gap.
Nevertheless, the transverse component of the SN interaction 
with a form of $I_i^+I_j^-$
is always absent in the field gradient because of the detuning effect, 
i.e., a mismatch in the Zeeman energies prevents the nuclei from exchanging magnons.
On the other hand, the longitudinal component ($I_i^zI_j^z$) of the SN interaction 
can survive even at low temperatures,
which is characterized by the antiferromagnetic spin-spin correlation function 
in the ground state (see \S \ref{sec:nuclear-interaction})
\cite{itoh96, sandvik96, kishine97}. 
This interaction, however, is short-ranged
because the spin-spin correlation function decays exponentially as a function of the distance,
so that the interaction can reach only a few lattice points
\cite{white94}.
As a result, the interactions between nuclei apart from each other by the order of ten lattice points 
are completely switched-off at low temperatures.

In order for the system to work as a QC, one has to provide logic gates.
A QC is complete if it is equipped with arbitrary rotation (R) and controlled-NOT (c-NOT) gates.
\cite{barenco95} 
The R gate is a single qubit operation, 
which can be accomplished by an NMR pulse with an appropriate pulse width at the corresponding frequency.
On the other hand, the c-NOT gate is a two-qubit operation, which works in such a way 
that a target qubit changes its logic according to the state of the control qubit, 
which can be realized by the pulse sequence shown in Fig. \ref{CNOT-gate}.
Since the c-NOT gate between any combination of qubits apart from each other along the chain 
can be realized by a series of c-NOT gates between {\it adjacent} qubits
\cite{collins00},
the QC is complete 
with the inter-nuclear coupling with the form $I_{iz}I_{jz}$ between adjacent nuclear spins.
This coupling is provided by the longitudinal component of the SN interaction 
via the $k=0$ magnon mode composed of the spin triplet state
\cite{zevin75}, 
which is selectively excited at a specific part of the chain 
where the c-NOT gate is to be performed.

\begin{figure}
\includegraphics[scale=0.3]{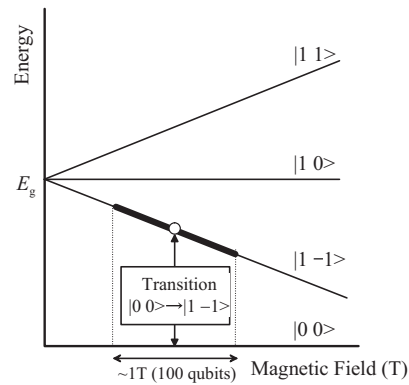}% Here is how to import EPS art
\caption{Energy diagrams for the three triplet branches 
($\mid 1 1 \rangle, \mid 1 0 \rangle$ and $\mid 1 -1 \rangle$) 
of the $k=0$ magnon excitation modes 
against the singlet state ($\mid 0 0 \rangle$) as a function of the external field.
In the field gradient, the horizontal axis also corresponds to the position along the chain.
The thick line on the $\mid1-1\rangle$ branch shows a part of the chain used as a QC, 
and an open circle is the spot 
where the transition ($\mid 0 0 \rangle \rightarrow \mid 1 -1 \rangle$) occurs.}
\label{microwave}
\end{figure}
The transition between singlet and triplet states by a microwave irradiation 
can be used to create the triplet $k=0$ magnons.
Although the excitation is primarily forbidden for the usual electric dipolar transition, 
it often becomes possible in the actual systems because of some higher order terms 
in the electron-photon interaction Hamiltonians
\cite{brill94,lu91}.
The position of the excited magnons along the chain 
can be specified by the applied microwave frequency, 
which is uniquely given in the field gradient.
The energy diagrams of the $k=0$ magnon excitations in the magnetic field is shown in Fig. \ref{microwave}
\cite{tachiki70, brill94}.
In the field gradient, the magnetic field at each part of the chain is unique 
so that the excitation energy to the lowest triplet state ($\mid1-1\rangle$) is also uniquely given,
which provides us with a spatial resolution of the excitation, i.e.,
one can specify the position of the magnon excitations by adjusting the frequency of the applied microwave.
The region and the amplitude of the excited magnons
can be controlled by the frequency resolution and the power of the microwave. 

Since the two upper states ($\mid1 0\rangle$, $\mid1 1\rangle$)
irrelevant to the transition of interest can be ignored, 
one can introduce effective {\it spins}
within the $\mid 0 0\rangle$ and $\mid 0 1\rangle$ subspace
\cite{tachiki70}. 
These {\it spins} are rotated by the microwave,
creating a packet of superpositions of $\mid 0 0\rangle$ and $\mid1 -1\rangle$ 
along the chain
\cite{brill94},
corresponding to the magnon excitations with the wave number $k \sim 0$.
The packet is localized along the chain due to the magnetic field gradient
(see \S\ref{sec:singlet-triplet}).
The number of excited {\it spins} is given by the balance 
between the excitation to the $\mid 1 -1\rangle$ state and the relaxation 
(with the lifetime of $T_s$) to the ground state.

The triplet states make an additional field ($H_{\rm tr}$) associated with the ``shift''
at the qubit site,
which should be distinguished from the additional field from the adjacent qubit 
via the SN interaction ($H_{\rm SN}$).
Assuming the hyperfine coupling $A_{\parallel}$ = 100 kOe/$\mu_B$ 
and the rotation of the {\it spins}: $n(0) \sim$ 1 \% ,
$H_{\rm tr}$ is estimated to be $\sim$ 1 kOe, 
which corresponds to $\sim$ 1 \% shift of the NMR frequency in the magnetic field of 10 T.
In practice, the exact value of $H_{\rm tr}$ can be measured by the following method; 
under the microwave irradiation, 
one observes the shift of the NMR frequency of the target qubit 
while saturating the control qubit by applying the corresponding NMR rf field continuously.
The saturation of the control qubit results in $H_{\rm SN}=0$ at the target qubit, 
so that the observed shift directly corresponds to $H_{\rm tr}$.

Then, the c-NOT gate is achieved as follows (see Fig. \ref{CNOT-gate}).
(a) In the beginning, the target qubit ($I_{N+1}$) is assumed to be pointing to the Z direction.
The $\pi/2$ pulse with the rf frequency $\omega=\gamma_n H(x_{N+1})$
is applied in the negative X direction in the rotating frame of the target qubit.
(b) A microwave is applied to the chain. 
The SN interaction is switched on, 
and the target qubit starts to rotate in the XY plane of the rotating frame 
with $\omega=\gamma_n(H(x_{N+1})+H_{tr})$.
The direction of the rotation with respect to the frame 
depends on the state of the control qubit ($I_N$).
(c) By the time $t=\pi/(2\gamma_n H_{\rm SN})$, 
the target qubit reaches either positive or negative X.
(d) The SN interaction is shut off by shutting off the microwave,
and the second $\pi/2$ pulse with $\gamma_n H(x_{N+1})$ 
is applied in the negative Y direction, 
which rotates the target qubit upward or downward according to the control qubit.
This sequence allows us to perform the c-NOT gate.

\section{Singlet-triplet excitations in the quantum spin chain}
\label{sec:singlet-triplet}
\begin{figure}[t]
\includegraphics[scale=0.6]{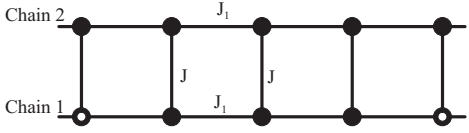}% Here is how to import EPS art
\caption{A spin ladder configuration. 
Closed and open circles represent electron and nuclear spins, respectively.}
\label{ladder}
\end{figure}

One of the key phenomena in the present model 
is the selective excitations of the triplet states
shown in Fig. ~\ref{microwave}
\cite{brill94}.
Let us see the excitation in detail.
We consider the ladder case shown in Fig. \ref{ladder} as an example, 
and follow the description in Ref. \onlinecite{tachiki70}. 
The effective {\it spin} is introduced within the subspace consisting
of the ground state $\mid0 0\rangle$ and the lowest excited state $\mid 1 -1\rangle$.
This treatment is helpful to visualize the transition as a rotation of the {\it spin}.
The Hamiltonians which govern the system are, 
\begin{eqnarray}
 H    &=& H_0+H_1, \label{H1}\\
 H_0 &=& J\sum_i {\bf s}_{i1} \cdot {\bf s}_{i2} + g\mu_B \sum_i {\bf H}(x_i) \cdot  ({\bf s}_{i1} + {\bf s}_{i2}),\label{H2}\\
 H_1 &=& J_1\sum_{nn}{\bf s}_{i1} \cdot {\bf s}_{j1}+J_1\sum_{nn}{\bf s}_{i2} \cdot {\bf s}_{j2},\label{H3}
\end{eqnarray}
Here, $J$ and $J_1$ are the intra- (rung) and inter-pair (leg) exchange interactions, respectively.
${\bf H}(x_i)$ is the external field at the $i$-th site and ``$nn$'' means the nearest neighbor sites.
By introducing the operators ${\textbf S}_i={\textbf s}_{i1} + {\textbf s}_{i2}$ and 
${\textbf T}_i={\textbf s}_{i1} - {\textbf s}_{i2}$,
Eqs. (\ref{H1})$\sim$(\ref{H3}) are rewritten as,
\begin{eqnarray}
 H_0 &=& \frac{1}{2}J\sum_i {\textbf S}_{i}^2 + g\mu_B \sum_i {\bf H}(x_{i}) \cdot {\bf S}_{i}\\
 H_1 &=& \frac{1}{2}J_1\sum_{nn} ({\textbf S}_{i} \cdot {\textbf S}_{j}+{\textbf T}_{i} \cdot {\textbf T}_{j}),
\end{eqnarray}
where some constant terms are dropped.

We further rewrite these Hamiltonians in terms of the Pauli spin components in the following manner.
\begin{eqnarray}
{\bf S}_i^2 =1-\sigma^z_{i}, & S^x_{i}=S^x_{j}=0, & S^z_{i}=\frac{1}{2}(\sigma^z_{i}-1), \nonumber\\
T^x_{i}=\frac{1}{\sqrt{2}} \sigma^x_{i}, & T^y_{i}=\frac{1}{\sqrt{2}} \sigma^y_{i}, &T^z_{i}=0. 
\label{spin}
\end{eqnarray}
As a consequence of these transformations, the two states of $\mid 0 0\rangle$ and $\mid 1 -1\rangle$ can be 
treated as the two spin states represented by the Pauli spin matrices.
The total Hamiltonian, Eq. (\ref{H1}) is expressed by these spin matrices as,
\begin{eqnarray}
H&=&\frac{N}{2}(J+\frac{J_1}{4}-g\mu_B \sum_i H(x_i)) \nonumber\\
 &-&\frac{1}{2}\sum_i \{J+\frac{J_1}{4}-g\mu_B H(x_i) \} \sigma_{zi} \nonumber\\
 &+&\frac{J_1}{8}\sum_{nn}\{2(\sigma^x_{i}\sigma^x_{j}+\sigma^y_{i}\sigma^y_{j})+\sigma^z_{i}\sigma^z_{j}\}
\end{eqnarray}
This describes the system of the {\it spins} coupled through anisotropic exchange interactions.
The microwave rotates these {\it spins}.

The small-angle rotations of the {\it spins} in a small region
create a packet of $k \sim 0$ magnons like a soliton. 
The packet is localized on the chain due to the magnetic field gradient.
A mismatch in the magnon excitation energies between adjacent regions along the chain 
prohibits the packet from moving to the lower field region. 
On the other hand, 
the continuum excitations near the one-magnon excitations at $k$=0 is absent,
so that it is difficult for the packet to move to the higher field region
\cite{augier98, uhrig96, garrett97, yu00, grenier00}
unless the process of the energy release by phonon emissions is considerable.
Consequently, the magnons are {\it confined} in the region where they are excited, 
and the SN interaction is produced only between the qubit pair of interest. 

The population of the magnons with $k=0$ ($\equiv n(0)$) is determined 
by the balance between excitation and relaxation,
\begin{equation}
 \frac{d n(0)}{dt}=W_{\rm ex}-\frac{n(0)}{T_s},
\end{equation}
where $W_{\rm ex}$ is the transition probability of $\mid 00 \rangle \rightarrow \mid 1 -1 \rangle$ 
per unit time by the microwave irradiation, and $T_s$ is the magnon lifetime.
At the steady state, $d n(0)/dt=0$, so that $n(0) = W_{\rm ex} T_s$.
As the microwave irradiation is shut off, the {\it spins} start to relax 
to the ground state ($\sigma_z=-1$) with the relaxation time given by $T_s$. 

We end up this section with some remarks on the nature of the excited states.
One is about the lifetime of the excited states ($T_s$);
it could be rather long 
because the excitation is primarily forbidden for the usual electric dipolar transition. 
Although the forbiddance itself is usually lifted by some additional interactions 
such as the Dzyaloshinsky-Moriya interaction
\cite{brill94,lu91},
the primarily forbidden transition is expected to lead to rather long lifetime 
of the excited triplet states and narrower transition line, 
which is favorable for the selective excitation.
Another is about the wave number of the triplet states;
they should have $k$=0 and 
the transition to the staggered component ($k=a/\pi$) is forbidden,
because only the transitions with the momentum transfer 
$q \equiv k-k^{'}=0$ are allowed by the microwave, and
the ground state ($|00 \rangle$) is a uniform singlet state ($k$=0).
This fact is quite favorable for the long-range inter-nuclear coupling with $q \sim 0$. 
Note that the magnons excited by the microwave are in the non-equilibrium states
far from the thermal equilibrium
where the $k=a/\pi$ magnons are primarily excited.

\section{Inter-nuclear couplings mediated by magnons}
\label{sec:nuclear-interaction}
We next look into the details of the longitudinal component of the SN interaction 
caused by a packet of spin triplet states.
For simplicity, we assume the following on-site anisotropic hyperfine Hamiltonian in the $i$-th site,
\begin{equation}
 H_{hf}=\{A_{\parallel}s_{i1}^zI_{i}^z + \frac{1}{2}A_{\perp} (s_{i1}^+I_{i}^- + s_{i1}^-I_{i}^+) \},
\end{equation}
which can be rewritten using the {\it spin} introduced in Eq. (\ref{spin}) as,
\begin{equation}
H_{hf} = \frac{1}{4} A_{\parallel}( I_i^z\sigma_i^z-I_i^z) + \frac{1}{2{\sqrt 2}} A_{\perp} (\sigma_i^+I_i^- +\sigma_i^-I_i^+). 
\end{equation}
Hence, besides the term, $-\frac{1}{4}A_{\parallel} I_i^z$, 
which can be incorporated into the Zeeman term in the nuclear Hamiltonian, 
the nuclear interaction with the {\it spin} $\sigma$ can be expressed by the anisotropic hyperfine interaction.
The first term creates the shifts at the nuclear sites corresponding to $H_{tr}$ and the SN interactions, 
and the other terms give rise to the spin-lattice relaxation
\cite{beeman68, pieper93, ivanov99}.
The effect of the spin lattice relaxation is discussed in the next section.

Since the transverse component of the SN interaction 
due to $\sigma^{\pm}$ vanishes in the field gradient,
we can restrict ourselves to the longitudinal component.
The longitudinal component of the SN interaction is given by,
\begin{equation}
 H_{SN} = W_{ij}I_i^zI_j^z,
\label{WII}
\end{equation}
where,
\begin{equation}
 W_{ij} = \left (\frac{\gamma_nA_{\parallel}}{N} \right )^2
\sum_{k,k^{'}, k \neq k^{'}} \frac{n_k-n_{k^{'}}}{\epsilon_{k^{'}}-\epsilon_k} \cos \{(k-k^{'}) r_{ij}\}.
\label{longitudinal-SN}
\end{equation}
Here, $n_k$ and $\epsilon_k$ are, respectively, the population and the energy 
of the magnon with the wave number $k$,
and $r_{ij}$ is the distance between the two nuclei of interest.
In the equilibrium states, 
$n_k$ is given by the Bose function for the given temperature.

Actually, eq. (\ref{WII}) is a special case of the general formulas 
for the indirect spin-spin interaction,
\begin{equation}
 H_{ind}=\Phi(r_{ij})I_i^zI_j^z,
\end{equation}
with the range function,
\begin{equation}
 \Phi(r_{ij})=\gamma_n^2 A_{\parallel}^2 \sum_{q} \chi (q) \exp (i q r_{ij}).
\end{equation}
Here, $\chi(q)$ is the zero energy component of the generalized susceptibility, 
and $q=k-k^{'}$.
In the present case, the electronic state is not in the equilibrium state,
so that $\chi(q)$ is different from that in the thermal equilibrium.
Note in particular, that the magnons with the wave numbers other than $k \sim 0$,
such as $k=\pi/a$, are not excited here.

The range to which this interaction reaches depends on the range function, 
which is determined by the form of $\chi(q)$ as a function of $q$.
For the pair of nuclei far from each other, 
the most important interaction comes from the uniform part of the susceptibility, i.e. 
$q \sim 0$ caused by the scatterings of the magnons with $k \sim 0$. 

In the case of the transverse component of the SN interaction in a 3-D system,
the range function can be calculated only from the magnon dispersions
and has the form $\sim a/r \exp(-r/3a)$ ($a$ is a lattice constant)
\cite{nakamura58}.
In our case, however, the situation is completely different from this case, 
and the interaction reaches rather long distance
because of the following reasons.
Firstly, in the longitudinal component, 
the number of excited magnons, as well as the magnon dispersions, 
are responsible for the range function (see Eq. (\ref{longitudinal-SN})).
Secondly, the system is 1-D 
so that a qualitative difference exists in the structures of $\chi(q)$.
Thirdly, the system is not in the equilibrium state.
Since $\chi(q)$ is enhanced at $q \sim 0$ in the present case, 
a rather long distance interaction is expected between the nuclei of interest.
Note again that the contribution from the magnons with $k \sim \pi/a$ to $\chi(q \sim 0)$,
which exists in the equilibrium state, does not exist in the present case.

Here, we make a rough estimation of the range function for the case of Fig. \ref{ladder}
using Eq. (\ref{longitudinal-SN}).
The magnon dispersion of the triplet state in this situation is given by
\cite{muller00},
\begin{eqnarray}
\epsilon(k_n) & = &C+J(j_1-\frac{1}{4}j_2^3) \cos(k_n)+ ...
\label{dispersion}
\end{eqnarray}
where $k_n  = n\pi /N$, $j_1 = J_1/J$ and $C$ is the part independent of $k_n$.
Here, the magnetic field is assumed to be uniform in this region. 
Using Eq. (\ref{dispersion}) and 
recalling that the microwave irradiation excites only the $k=0$ component of the magnons, 
i.e., $n(k)=0$ for $k \neq 0$, 
one can calculate the range function $W_{ij}$ in Eq. (\ref{longitudinal-SN}),
\begin{eqnarray}
 W_{ij} & = & \left (\frac{\gamma_n A_{\parallel}}{N} \right )^2
\sum_{n=1}^{N} \frac{2 n(0)}{\epsilon(k_n)-\epsilon(0)} \cos (k_n r_{ij}) \nonumber\\
 & = & \frac{2 \gamma_n^2 A_{\parallel}^2 \{n(0)/N\} }{J(j_1-\frac{1}{4}j_1^3)N}
 \sum_{n=1}^N \frac{\cos (k_n r_{ij})}{\cos (k_n)-1}.
\end{eqnarray}
Assuming $N$ = 20, $r_{ij}$ = 10 (in units of $a$), $A_{\parallel}$ = 100 kOe/$\mu_B$, 
$\gamma_n/(2 \pi)$ = 4.3 MHz/kOe ($^{1}$H as an example), $J$ = 50 K, $j_2$ = 0.2 
and $n(0)/N = 0.01$,
one obtains $W_{ij}$ = 15 kHz,
which is the same order of magnitude as
the nuclear dipole coupling acting between nuclei 3 \AA \hspace{1mm} apart from each other
\cite{yamaguchi99, ladd00}, 
and one to three orders of magnitude greater than the J-couplings used in the solution NMR-QC's
\cite{chuang98, jones98}.

The estimation here is for the ideal case, 
because the population difference may be underestimated
($n(k \neq 0)$ may be populated somewhat in the actual situation).
Moreover, the realistic magnon dispersions
as well as the conditions of the microwave excitation such as $n(0)/N$ and $N$
may affect the value of $W_{ij}$.
Nevertheless, this estimation indicates that 
the longitudinal component of the SN interaction could be large enough to serve as a c-NOT gate. 
Note that the strength of the coupling, 
$W_{ij}$ can be controlled by the microwave intensity via $n(0)$,
because it is determined by the balance between excitation and relaxation, 
and in the steady state, $n(0) = W_{\rm ex} T_s$.

\section{Decoherence}
\label{sec:decoherence}
Decoherence is one of the serious concerns in this model.
On the one hand, the excited triplet states can create 
inter-qubit couplings, but on the other hand, 
they inevitably activate the scattering channels of the nuclear spins
and reduce the spin-lattice relaxation time $T_1$
\cite{beeman68, pieper93, ivanov99}.
Although this is unavoidable, it is still possible 
to reduce the chances of decoherence.

Let us see the scattering process of magnons by a nucleus in detail.
Since the direct one-magnon process is prohibited 
because of the mismatch between nuclear and electron excitation energies,
the possible lowest order process is the two-magnon (Raman) process.
This process can be the major cause of $T_1$ relaxation
when the number of excited magnons is increased. 
The process is known, however, to require highly anisotropic exchange interactions
because of the conservation of the angular momentum before and after the scattering.
Hence, the process can be reduced in the system with isotropic exchange interactions 
\cite{beeman68, pieper93}.
The three-magnon process is the next possible process, where two incident magnons are scattered 
by a nucleus and one magnon is emitted. 
This process is active even in the case of isotropic exchange interactions.

In general, the spin-lattice relaxation time is given by,
\begin{equation}
  \frac{1}{T_1}= \frac{2 \gamma_n^2 A_{\perp}^2 k_B T}{g^2 \mu_B^2} \sum_q \frac{{\rm Im}\chi^{+-} (q, \omega_n)}{\omega_n}.
\label{T1}
\end{equation}
here, $\chi^{+-}(q,\omega)$ is the transverse component of the dynamical susceptibility
and $\omega_n$ is the NMR frequency.
(Note again, that the present situation is not a thermal equilibrium state,
so that $\chi(q,\omega)$ is different from those in the thermal equilibrium.)
As seen in Eq. (\ref{T1}), $1/T_1$ is proportional to $A_{\perp}^2$.
Recalling that $A_{\parallel}$ is the only necessary component of the hyperfine coupling 
for the longitudinal component of the SN interaction, 
the highly anisotropic hyperfine coupling ($A_{\parallel}/A_{\perp} \gg 1$)
may be used to reduce $1/T_1$ without reducing the longitudinal component of the SN interaction. 
Such highly anisotropic hyperfine interaction can be realized 
by transfered hyperfine couplings
\cite{mila89}. 

Even with such effects, however, 
it might be inevitable to invoke the quantum error correction method 
for the unavoidable nuclear spin flips in the end
\cite{shor95, steane96}.
Even so, the chain structure is fortunate for the error correction process.
The error correction requires three equivalent QC's,
which are entangled like $\alpha_{l1}\alpha_{l2}\alpha_{l3}$ at each qubit $l$, 
where $\alpha_{li}$ is an eigenstate (e.g. spin-up state) at the qubit $l$ of the i-th QC.
Such set of nuclei can be provided by the three consecutive chains.
It is possible to align the qubit nuclei (see \S \ref{sec:alignments}) and once they are aligned,
the inter-chain nuclear dipole couplings keep them entangled.
(For this purpose, only the $I^zI^z$ component is needed.
Since $I^+I^-$ components make the NMR line broad, 
they should be eliminated by applying a small field gradient in this direction.)

\section{Nuclear alignments}
\label{sec:alignments}
So far, we have described the model for a {\it single} QC.
Unfortunately, the sensitivity of the conventional NMR method
is not high enough to detect signals from a single QC, 
so that one needs to integrate many equivalent QC's to obtain the results of computations.
In this process, the {\it nuclear alignment} can help us to reduce 
the number of equivalent nuclei required for each qubit.
It is also advantageous for obtaining a narrower NMR line.

The nuclear alignment is also associated with the qubit initialization.
Although the qubit initialization may be achieved by the pseudo-pure state technique
\cite{gershenfeld97}
and/or the algorithmic cooling
\cite{schulman99,ladd02},
aligning nuclear spins, even partially, 
can assist the methods to work in the systems with the large number of qubits ($N$),
because the number of QC's which happen to be in the pure state in the thermal equilibrium
is proportional to $(\hbar \omega_n/2k_B T)N/2^N$,
which becomes smaller and smaller as increasing $N$. 
The nuclear alignment can relieve this problem.

\begin{figure}[t]
\includegraphics[scale=0.38]{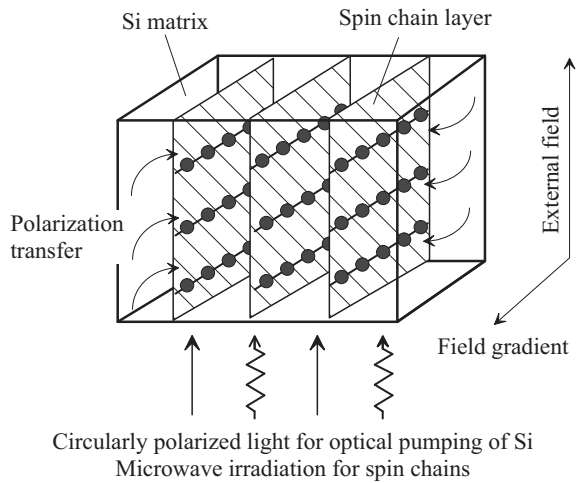}% Here is how to import EPS art
\caption{Schematic illustration of a {\it qubit initializer} for
an integrated NMR-QC and a possible experimental configuration.
Films made of spin chains are embedded in a Si matrix
and placed in a field gradient and at a low temperature.
Small balls represent the qubit nuclei in the spin chains.
A laser light for the optical pumping of the electrons in the Si matrix and 
the microwave irradiation for the spin chains are delivered from the bottom
(Faraday configuration).
}
\label{Si-matrix}
\end{figure}
Here, we propose a possible configuration of a ``qubit initializer'',
which partially aligns the qubits in the spin chains by the optical pumping technique 
\cite{tycko98}.
The scheme enables us to separate the materials responsible 
for the initialization and computation, 
so that one can optimize the computation part
without being restricted by the initialization scheme. 
The schematic illustration is shown in Fig. \ref{Si-matrix}. 
Films made of spin chains are sandwiched by Si single crystals
and placed in a field gradient and at a low temperature. 
The electrons in the semiconducting Si matrix are polarized
by a circularly polarized near-infrared laser light 
with an energy corresponding to the semiconducting gap of Si
\cite{lampel68, meier84}, 
and the polarizations of the electrons are transfered 
to the nuclear spin systems of the spin chains 
through the hyperfine couplings
\cite{michal98, michal99, farah98}, 
or the cross-polarization /coherent transfer techniques
\cite{slichter90}. 
Here, the films and the Si matrix are not necessarily bonded chemically
but contacted mechanically,
because the direct dipolar couplings are available for the polarization transfer process
\cite{tycko98}.

After the nuclear alignment is completed, the laser light is shut off.
The excited electrons in the Si matrix will die out quickly
and the Si matrix will return to the silent environment. 
Here, the Si matrix is selected 
because the abundant nuclei ($^{28}$Si) have no spins ($I = 0$).
Since the optical pumping is effective in most of the semiconductors
\cite{meier84},
other semiconductors with no nuclear spins can be used as well.
For example, 
CdTe enriched by the isotopes with no nuclear spins, 
such as $^{110,112,114}$Cd and $^{126,128,130}$Te is also available.

\section{Practical issues in the implementation}
\label{sec:implementation}
We have shown an ideal model for the NMR QC so far,
but one might notice some practical issues to implement this model.
First of all, one should find a suitable spin chain system with a singlet ground state, 
containing more than two stable isotopes for one element, 
one of which has $I=1/2$.
Moreover, the periodical placements of the 1/2 nuclear spins may require some ingenious techniques.

One of the possible candidates for the QC 
would be a film made of some organic materials
with spin gaps.
They contain $^1$H and $^{13}$C having $I=1/2$.
They also have large unit cells, 
which are useful to reserve large spacial distances between qubits.
Moreover, one could utilize a well-established chemical technique of 
the selective isotope replacement of $^1$H by $^2$D ($I=1$) 
\cite{aonuma93a}
or $^{12}$C ($I=0$) by $^{13}$C ($I=1/2$).
The periodic placements of the qubits may be achieved 
with the epitaxial growth
\cite{schuerlein95}
and/or the Langmuir-Blodgett methods as follows.
One prepares two sets of molecules with the same chemical formulas but different isotopes 
by the selective isotope replacement technique, 
for example, all deuterized samples and those having $^1$H's at one of the hydrogen sites.
These two sets of molecules could be layered epitaxially using the MBE 
or the Langmuir-Blodgett method
so as for $^1$H's or $^{13}$C's to be placed periodically.

In order to avoid unnecessary nuclear couplings,
the nuclei other than qubits are preferable to have no spins ($I=0$).
In this sense, it is fortunate that major abundant isotopes 
in organic materials such as $^{12}$C and $^{16}$O have no spins.
The Haldane systems such as NENP and NINO are also fortunate
because they contain $^{58}$Ni, an abundant isotope with no spins.
In addition, a high power decoupler and/or 
a decoupling sequence between unlike-spins using a train of $\pi$-pulses
can counteract the effects of the nuclei with $I \neq 0$ such as $^2$D.

\begin{acknowledgments}
We wish to acknowledge helpful advices by G. Kido and M. Kitagawa. 
This work has been supported by CREST of JST (Japan Science and Technology Corporation).
\end{acknowledgments}

\bibliography{j-switch}% Produces the bibliography via BibTeX.

\begin{thebibliography}{50}
\expandafter\ifx\csname natexlab\endcsname\relax\def\natexlab#1{#1}\fi
\expandafter\ifx\csname bibnamefont\endcsname\relax
  \def\bibnamefont#1{#1}\fi
\expandafter\ifx\csname bibfnamefont\endcsname\relax
  \def\bibfnamefont#1{#1}\fi
\expandafter\ifx\csname citenamefont\endcsname\relax
  \def\citenamefont#1{#1}\fi
\expandafter\ifx\csname url\endcsname\relax
  \def\url#1{\texttt{#1}}\fi
\expandafter\ifx\csname urlprefix\endcsname\relax\def\urlprefix{URL }\fi
\providecommand{\bibinfo}[2]{#2}
\providecommand{\eprint}[2][]{\url{#2}}

\bibitem[{\citenamefont{Deutsch}(1985)}]{deutsch85}
\bibinfo{author}{\bibfnamefont{D.}~\bibnamefont{Deutsch}},
  \bibinfo{journal}{Proc.\ R.\ Soc.\ Lond.\ A} \textbf{\bibinfo{volume}{400}},
  \bibinfo{pages}{97} (\bibinfo{year}{1985}).

\bibitem[{\citenamefont{Shor}(1997)}]{shor97}
\bibinfo{author}{\bibfnamefont{P.~W.} \bibnamefont{Shor}},
  \bibinfo{journal}{SIAM, J. Compt.} \textbf{\bibinfo{volume}{26}},
  \bibinfo{pages}{1417} (\bibinfo{year}{1997}).

\bibitem[{\citenamefont{Grover}(1998)}]{grover98}
\bibinfo{author}{\bibfnamefont{L.~K.} \bibnamefont{Grover}},
  \bibinfo{journal}{Phys.\ Rev.\ Lett.} \textbf{\bibinfo{volume}{80}},
  \bibinfo{pages}{4329} (\bibinfo{year}{1998}).

\bibitem[{\citenamefont{Chuang et~al.}(1998)\citenamefont{Chuang, Gershenfeld,
  and Kubinec}}]{chuang98}
\bibinfo{author}{\bibfnamefont{I.~L.} \bibnamefont{Chuang}},
  \bibinfo{author}{\bibfnamefont{N.}~\bibnamefont{Gershenfeld}},
  \bibnamefont{and} \bibinfo{author}{\bibfnamefont{M.}~\bibnamefont{Kubinec}},
  \bibinfo{journal}{Phys.\ Rev.\ Lett.} \textbf{\bibinfo{volume}{80}},
  \bibinfo{pages}{3408} (\bibinfo{year}{1998}).

\bibitem[{\citenamefont{Jones and Mosca}(1998)}]{jones98}
\bibinfo{author}{\bibfnamefont{J.~A.} \bibnamefont{Jones}} \bibnamefont{and}
  \bibinfo{author}{\bibfnamefont{M.}~\bibnamefont{Mosca}}, \bibinfo{journal}{J.
  Chem.\ Phys.} \textbf{\bibinfo{volume}{109}}, \bibinfo{pages}{1648}
  (\bibinfo{year}{1998}).

\bibitem[{\citenamefont{Kane}(1998)}]{kane98}
\bibinfo{author}{\bibfnamefont{B.~E.} \bibnamefont{Kane}},
  \bibinfo{journal}{Nature} \textbf{\bibinfo{volume}{393}},
  \bibinfo{pages}{133} (\bibinfo{year}{1998}).

\bibitem[{\citenamefont{Yamaguchi and Yamamoto}(1999)}]{yamaguchi99}
\bibinfo{author}{\bibfnamefont{F.}~\bibnamefont{Yamaguchi}} \bibnamefont{and}
  \bibinfo{author}{\bibfnamefont{Y.}~\bibnamefont{Yamamoto}},
  \bibinfo{journal}{Appl.\ Phys.\ A} \textbf{\bibinfo{volume}{68}},
  \bibinfo{pages}{1} (\bibinfo{year}{1999}).

\bibitem[{\citenamefont{Hashi et~al.}(2000)\citenamefont{Hashi, Shimizu, Goto,
  Kitazawa, Kido, and Suzuki}}]{hashi00}
\bibinfo{author}{\bibfnamefont{K.}~\bibnamefont{Hashi}},
  \bibinfo{author}{\bibfnamefont{T.}~\bibnamefont{Shimizu}},
  \bibinfo{author}{\bibfnamefont{A.}~\bibnamefont{Goto}},
  \bibinfo{author}{\bibfnamefont{H.}~\bibnamefont{Kitazawa}},
  \bibinfo{author}{\bibfnamefont{G.}~\bibnamefont{Kido}}, \bibnamefont{and}
  \bibinfo{author}{\bibfnamefont{T.}~\bibnamefont{Suzuki}},
  \bibinfo{journal}{Appl.\ Phys.\ A} \textbf{\bibinfo{volume}{70}},
  \bibinfo{pages}{359} (\bibinfo{year}{2000}).

\bibitem[{\citenamefont{Ladd et~al.}(2000)\citenamefont{Ladd, Goldman,
  Yamaguchi, and Yamamoto}}]{ladd00}
\bibinfo{author}{\bibfnamefont{T.~D.} \bibnamefont{Ladd}},
  \bibinfo{author}{\bibfnamefont{J.~R.} \bibnamefont{Goldman}},
  \bibinfo{author}{\bibfnamefont{F.}~\bibnamefont{Yamaguchi}},
  \bibnamefont{and} \bibinfo{author}{\bibfnamefont{Y.}~\bibnamefont{Yamamoto}},
  \bibinfo{journal}{Appl.\ Phys.\ A} \textbf{\bibinfo{volume}{71}},
  \bibinfo{pages}{27} (\bibinfo{year}{2000}).

\bibitem[{\citenamefont{Goldman et~al.}(2000)\citenamefont{Goldman, Ladd,
  Yamaguchi, and Yamamoto}}]{goldman00}
\bibinfo{author}{\bibfnamefont{J.~R.} \bibnamefont{Goldman}},
  \bibinfo{author}{\bibfnamefont{T.~D.} \bibnamefont{Ladd}},
  \bibinfo{author}{\bibfnamefont{F.}~\bibnamefont{Yamaguchi}},
  \bibnamefont{and} \bibinfo{author}{\bibfnamefont{Y.}~\bibnamefont{Yamamoto}},
  \bibinfo{journal}{Appl.\ Phys.\ A} \textbf{\bibinfo{volume}{71}},
  \bibinfo{pages}{11} (\bibinfo{year}{2000}).

\bibitem[{\citenamefont{Goto et~al.}(2002{\natexlab{a}})\citenamefont{Goto,
  Shimizu, Miyabe, Hashi, Kitazawa, Kido, Shimamura, and Fukuda}}]{goto02a}
\bibinfo{author}{\bibfnamefont{A.}~\bibnamefont{Goto}},
  \bibinfo{author}{\bibfnamefont{T.}~\bibnamefont{Shimizu}},
  \bibinfo{author}{\bibfnamefont{R.}~\bibnamefont{Miyabe}},
  \bibinfo{author}{\bibfnamefont{K.}~\bibnamefont{Hashi}},
  \bibinfo{author}{\bibfnamefont{H.}~\bibnamefont{Kitazawa}},
  \bibinfo{author}{\bibfnamefont{G.}~\bibnamefont{Kido}},
  \bibinfo{author}{\bibfnamefont{K.}~\bibnamefont{Shimamura}},
  \bibnamefont{and} \bibinfo{author}{\bibfnamefont{T.}~\bibnamefont{Fukuda}},
  \bibinfo{journal}{Appl.\ Phys.\ A} \textbf{\bibinfo{volume}{74}},
  \bibinfo{pages}{73} (\bibinfo{year}{2002}{\natexlab{a}}).

\bibitem[{\citenamefont{Ruderman and Kittel}(1954)}]{ruderman54}
\bibinfo{author}{\bibfnamefont{M.~A.} \bibnamefont{Ruderman}} \bibnamefont{and}
  \bibinfo{author}{\bibfnamefont{C.}~\bibnamefont{Kittel}},
  \bibinfo{journal}{Phys.\ Rev.} \textbf{\bibinfo{volume}{96}},
  \bibinfo{pages}{99} (\bibinfo{year}{1954}).

\bibitem[{\citenamefont{Suhl}(1958)}]{suhl58}
\bibinfo{author}{\bibfnamefont{H.}~\bibnamefont{Suhl}},
  \bibinfo{journal}{Phys.\ Rev.} \textbf{\bibinfo{volume}{109}},
  \bibinfo{pages}{606} (\bibinfo{year}{1958}).

\bibitem[{\citenamefont{Nakamura}(1958)}]{nakamura58}
\bibinfo{author}{\bibfnamefont{T.}~\bibnamefont{Nakamura}},
  \bibinfo{journal}{Prog.\ Theor.\ Phys.} \textbf{\bibinfo{volume}{20}},
  \bibinfo{pages}{542} (\bibinfo{year}{1958}).

\bibitem[{\citenamefont{Goto et~al.}(2002{\natexlab{b}})\citenamefont{Goto,
  Shimizu, and Hashi}}]{goto02b}
\bibinfo{author}{\bibfnamefont{A.}~\bibnamefont{Goto}},
  \bibinfo{author}{\bibfnamefont{T.}~\bibnamefont{Shimizu}}, \bibnamefont{and}
  \bibinfo{author}{\bibfnamefont{K.}~\bibnamefont{Hashi}}
  (\bibinfo{year}{2002}{\natexlab{b}}), \eprint{quant-ph/0201060}.

\bibitem[{\citenamefont{Khitun et~al.}(2001)\citenamefont{Khitun, Ostroumov,
  and Wang}}]{khitun01}
\bibinfo{author}{\bibfnamefont{A.}~\bibnamefont{Khitun}},
  \bibinfo{author}{\bibfnamefont{R.}~\bibnamefont{Ostroumov}},
  \bibnamefont{and} \bibinfo{author}{\bibfnamefont{K.~L.} \bibnamefont{Wang}},
  \bibinfo{journal}{Phys.\ Rev.\ A} \textbf{\bibinfo{volume}{64}},
  \bibinfo{pages}{62304} (\bibinfo{year}{2001}).

\bibitem[{\citenamefont{Itoh et~al.}(1996)\citenamefont{Itoh, Sugahara,
  Yamauchi, and Ueda}}]{itoh96}
\bibinfo{author}{\bibfnamefont{M.}~\bibnamefont{Itoh}},
  \bibinfo{author}{\bibfnamefont{M.}~\bibnamefont{Sugahara}},
  \bibinfo{author}{\bibfnamefont{T.}~\bibnamefont{Yamauchi}}, \bibnamefont{and}
  \bibinfo{author}{\bibfnamefont{Y.}~\bibnamefont{Ueda}},
  \bibinfo{journal}{Phys.\ Rev.\ B} \textbf{\bibinfo{volume}{54}},
  \bibinfo{pages}{R9631} (\bibinfo{year}{1996}).

\bibitem[{\citenamefont{Sandvik et~al.}(1996)\citenamefont{Sandvik, Dagotto,
  and Scalapino}}]{sandvik96}
\bibinfo{author}{\bibfnamefont{A.~W.} \bibnamefont{Sandvik}},
  \bibinfo{author}{\bibfnamefont{E.}~\bibnamefont{Dagotto}}, \bibnamefont{and}
  \bibinfo{author}{\bibfnamefont{D.~J.} \bibnamefont{Scalapino}},
  \bibinfo{journal}{Phys.\ Rev.\ B.} \textbf{\bibinfo{volume}{53}},
  \bibinfo{pages}{R2934} (\bibinfo{year}{1996}).

\bibitem[{\citenamefont{Kishine}(1997)}]{kishine97}
\bibinfo{author}{\bibfnamefont{J.}~\bibnamefont{Kishine}},
  \bibinfo{journal}{J.\ Phys.\ Soc.\ Jpn.} \textbf{\bibinfo{volume}{66}},
  \bibinfo{pages}{1229} (\bibinfo{year}{1997}).

\bibitem[{\citenamefont{White et~al.}(1994)\citenamefont{White, Noack, and
  Scalapino}}]{white94}
\bibinfo{author}{\bibfnamefont{S.~R.} \bibnamefont{White}},
  \bibinfo{author}{\bibfnamefont{R.~M.} \bibnamefont{Noack}}, \bibnamefont{and}
  \bibinfo{author}{\bibfnamefont{D.~J.} \bibnamefont{Scalapino}},
  \bibinfo{journal}{Phys.\ Rev.\ Lett.} \textbf{\bibinfo{volume}{73}},
  \bibinfo{pages}{886} (\bibinfo{year}{1994}).

\bibitem[{\citenamefont{Barenco et~al.}(1995)\citenamefont{Barenco, Bennett,
  Cleve, DiVincenzo, Margolus, Shor, Sleator, Smolin, and
  Weinfurter}}]{barenco95}
\bibinfo{author}{\bibfnamefont{A.}~\bibnamefont{Barenco}},
  \bibinfo{author}{\bibfnamefont{C.~H.} \bibnamefont{Bennett}},
  \bibinfo{author}{\bibfnamefont{R.}~\bibnamefont{Cleve}},
  \bibinfo{author}{\bibfnamefont{D.~P.} \bibnamefont{DiVincenzo}},
  \bibinfo{author}{\bibfnamefont{N.}~\bibnamefont{Margolus}},
  \bibinfo{author}{\bibfnamefont{P.}~\bibnamefont{Shor}},
  \bibinfo{author}{\bibfnamefont{T.}~\bibnamefont{Sleator}},
  \bibinfo{author}{\bibfnamefont{J.~A.} \bibnamefont{Smolin}},
  \bibnamefont{and}
  \bibinfo{author}{\bibfnamefont{H.}~\bibnamefont{Weinfurter}},
  \bibinfo{journal}{Phys.\ Rev.\ A} \textbf{\bibinfo{volume}{52}},
  \bibinfo{pages}{3457} (\bibinfo{year}{1995}).

\bibitem[{\citenamefont{Collins et~al.}(2000)\citenamefont{Collins, Kim,
  Holton, Sierzputowska-Gracz, and Stejskal}}]{collins00}
\bibinfo{author}{\bibfnamefont{D.}~\bibnamefont{Collins}},
  \bibinfo{author}{\bibfnamefont{K.~W.} \bibnamefont{Kim}},
  \bibinfo{author}{\bibfnamefont{W.~C.} \bibnamefont{Holton}},
  \bibinfo{author}{\bibfnamefont{H.}~\bibnamefont{Sierzputowska-Gracz}},
  \bibnamefont{and} \bibinfo{author}{\bibfnamefont{E.~O.}
  \bibnamefont{Stejskal}}, \bibinfo{journal}{Phys.\ Rev.\ A}
  \textbf{\bibinfo{volume}{62}}, \bibinfo{pages}{22304} (\bibinfo{year}{2000}).

\bibitem[{\citenamefont{Zevin and Kaplan}(1975)}]{zevin75}
\bibinfo{author}{\bibfnamefont{V.}~\bibnamefont{Zevin}} \bibnamefont{and}
  \bibinfo{author}{\bibfnamefont{N.}~\bibnamefont{Kaplan}},
  \bibinfo{journal}{Phys.\ Rev.\ B} \textbf{\bibinfo{volume}{12}},
  \bibinfo{pages}{4604} (\bibinfo{year}{1975}).

\bibitem[{\citenamefont{Brill et~al.}(1994)\citenamefont{Brill, Boucher,
  Voiron, Dhalenne, Revcolevschi, and Renard}}]{brill94}
\bibinfo{author}{\bibfnamefont{T.~M.} \bibnamefont{Brill}},
  \bibinfo{author}{\bibfnamefont{J.~P.} \bibnamefont{Boucher}},
  \bibinfo{author}{\bibfnamefont{J.}~\bibnamefont{Voiron}},
  \bibinfo{author}{\bibfnamefont{G.}~\bibnamefont{Dhalenne}},
  \bibinfo{author}{\bibfnamefont{A.}~\bibnamefont{Revcolevschi}},
  \bibnamefont{and} \bibinfo{author}{\bibfnamefont{J.~P.}
  \bibnamefont{Renard}}, \bibinfo{journal}{Phys.\ Rev.\ Lett.}
  \textbf{\bibinfo{volume}{73}}, \bibinfo{pages}{1545} (\bibinfo{year}{1994}).

\bibitem[{\citenamefont{Lu et~al.}(1991)\citenamefont{Lu, Tuchendler, von
  Ortenberg, and Renard}}]{lu91}
\bibinfo{author}{\bibfnamefont{W.}~\bibnamefont{Lu}},
  \bibinfo{author}{\bibfnamefont{J.}~\bibnamefont{Tuchendler}},
  \bibinfo{author}{\bibfnamefont{M.}~\bibnamefont{von Ortenberg}},
  \bibnamefont{and} \bibinfo{author}{\bibfnamefont{J.~P.}
  \bibnamefont{Renard}}, \bibinfo{journal}{Phys.\ Rev.\ Lett.}
  \textbf{\bibinfo{volume}{67}}, \bibinfo{pages}{3716} (\bibinfo{year}{1991}).

\bibitem[{\citenamefont{Tachiki and Yamada}(1970)}]{tachiki70}
\bibinfo{author}{\bibfnamefont{M.}~\bibnamefont{Tachiki}} \bibnamefont{and}
  \bibinfo{author}{\bibfnamefont{T.}~\bibnamefont{Yamada}},
  \bibinfo{journal}{J. Phys.\ Soc.\ Jpn.} \textbf{\bibinfo{volume}{28}},
  \bibinfo{pages}{1413} (\bibinfo{year}{1970}).

\bibitem[{\citenamefont{Augier and Poilblanc}(1998)}]{augier98}
\bibinfo{author}{\bibfnamefont{D.}~\bibnamefont{Augier}} \bibnamefont{and}
  \bibinfo{author}{\bibfnamefont{D.}~\bibnamefont{Poilblanc}},
  \bibinfo{journal}{Eur.\ Phys.\ J.\ B} \textbf{\bibinfo{volume}{1}},
  \bibinfo{pages}{19} (\bibinfo{year}{1998}).

\bibitem[{\citenamefont{Uhrig and Schulz}(1996)}]{uhrig96}
\bibinfo{author}{\bibfnamefont{G.~S.} \bibnamefont{Uhrig}} \bibnamefont{and}
  \bibinfo{author}{\bibfnamefont{H.~J.} \bibnamefont{Schulz}},
  \bibinfo{journal}{Phys.\ Rev.\ B} \textbf{\bibinfo{volume}{54}},
  \bibinfo{pages}{R9624} (\bibinfo{year}{1996}).

\bibitem[{\citenamefont{Garrett et~al.}(1997)\citenamefont{Garrett, Nagler,
  Tennant, Sales, and Barnes}}]{garrett97}
\bibinfo{author}{\bibfnamefont{A.~W.} \bibnamefont{Garrett}},
  \bibinfo{author}{\bibfnamefont{S.~E.} \bibnamefont{Nagler}},
  \bibinfo{author}{\bibfnamefont{D.~A.} \bibnamefont{Tennant}},
  \bibinfo{author}{\bibfnamefont{B.~C.} \bibnamefont{Sales}}, \bibnamefont{and}
  \bibinfo{author}{\bibfnamefont{T.}~\bibnamefont{Barnes}},
  \bibinfo{journal}{Phys.\ Rev.\ Lett.} \textbf{\bibinfo{volume}{79}},
  \bibinfo{pages}{745} (\bibinfo{year}{1997}).

\bibitem[{\citenamefont{Yu and Haas}(2000)}]{yu00}
\bibinfo{author}{\bibfnamefont{W.}~\bibnamefont{Yu}} \bibnamefont{and}
  \bibinfo{author}{\bibfnamefont{S.}~\bibnamefont{Haas}},
  \bibinfo{journal}{Phys.\ Rev.\ B} \textbf{\bibinfo{volume}{62}},
  \bibinfo{pages}{344} (\bibinfo{year}{2000}).

\bibitem[{\citenamefont{Grenier et~al.}(2000)\citenamefont{Grenier, Regnault,
  Lorenzo, Boucher, Hiess, Dhalenne, and Revcolevschi}}]{grenier00}
\bibinfo{author}{\bibfnamefont{B.}~\bibnamefont{Grenier}},
  \bibinfo{author}{\bibfnamefont{L.~P.} \bibnamefont{Regnault}},
  \bibinfo{author}{\bibfnamefont{J.~E.} \bibnamefont{Lorenzo}},
  \bibinfo{author}{\bibfnamefont{J.~P.} \bibnamefont{Boucher}},
  \bibinfo{author}{\bibfnamefont{A.}~\bibnamefont{Hiess}},
  \bibinfo{author}{\bibfnamefont{G.}~\bibnamefont{Dhalenne}}, \bibnamefont{and}
  \bibinfo{author}{\bibfnamefont{A.}~\bibnamefont{Revcolevschi}},
  \bibinfo{journal}{Phys.\ Rev.\ B} \textbf{\bibinfo{volume}{62}},
  \bibinfo{pages}{12206} (\bibinfo{year}{2000}).

\bibitem[{\citenamefont{Beeman and Pincus}(1968)}]{beeman68}
\bibinfo{author}{\bibfnamefont{D.}~\bibnamefont{Beeman}} \bibnamefont{and}
  \bibinfo{author}{\bibfnamefont{P.}~\bibnamefont{Pincus}},
  \bibinfo{journal}{Phys.\ Rev.} \textbf{\bibinfo{volume}{166}},
  \bibinfo{pages}{359} (\bibinfo{year}{1968}).

\bibitem[{\citenamefont{Pieper et~al.}(1993)\citenamefont{Pieper, Kotzler, and
  Nehrke}}]{pieper93}
\bibinfo{author}{\bibfnamefont{M.~W.} \bibnamefont{Pieper}},
  \bibinfo{author}{\bibfnamefont{J.}~\bibnamefont{Kotzler}}, \bibnamefont{and}
  \bibinfo{author}{\bibfnamefont{K.}~\bibnamefont{Nehrke}},
  \bibinfo{journal}{Phys.\ Rev.\ B} \textbf{\bibinfo{volume}{47}},
  \bibinfo{pages}{11962} (\bibinfo{year}{1993}).

\bibitem[{\citenamefont{Ivanov and Lee}(1999)}]{ivanov99}
\bibinfo{author}{\bibfnamefont{D.~A.} \bibnamefont{Ivanov}} \bibnamefont{and}
  \bibinfo{author}{\bibfnamefont{P.~A.} \bibnamefont{Lee}},
  \bibinfo{journal}{Phys.\ Rev.\ B} \textbf{\bibinfo{volume}{59}},
  \bibinfo{pages}{4803} (\bibinfo{year}{1999}).

\bibitem[{\citenamefont{Muller and Mikeska}(2000)}]{muller00}
\bibinfo{author}{\bibfnamefont{M.}~\bibnamefont{Muller}} \bibnamefont{and}
  \bibinfo{author}{\bibfnamefont{H.-J.} \bibnamefont{Mikeska}},
  \bibinfo{journal}{J. Phys.\ Condens.\ Matter} \textbf{\bibinfo{volume}{12}},
  \bibinfo{pages}{7633} (\bibinfo{year}{2000}).

\bibitem[{\citenamefont{Mila and Rice}(1989)}]{mila89}
\bibinfo{author}{\bibfnamefont{F.}~\bibnamefont{Mila}} \bibnamefont{and}
  \bibinfo{author}{\bibfnamefont{T.~M.} \bibnamefont{Rice}},
  \bibinfo{journal}{Physica C} \textbf{\bibinfo{volume}{157}},
  \bibinfo{pages}{561} (\bibinfo{year}{1989}).

\bibitem[{\citenamefont{Shor}(1995)}]{shor95}
\bibinfo{author}{\bibfnamefont{P.}~\bibnamefont{Shor}},
  \bibinfo{journal}{Phys.\ Rev.\ A} \textbf{\bibinfo{volume}{52}},
  \bibinfo{pages}{2493} (\bibinfo{year}{1995}).

\bibitem[{\citenamefont{Steane}(1996)}]{steane96}
\bibinfo{author}{\bibfnamefont{A.~M.} \bibnamefont{Steane}},
  \bibinfo{journal}{Phys.\ Rev.\ Lett.} \textbf{\bibinfo{volume}{77}},
  \bibinfo{pages}{793} (\bibinfo{year}{1996}).

\bibitem[{\citenamefont{Gershenfeld and Chuang}(1997)}]{gershenfeld97}
\bibinfo{author}{\bibfnamefont{N.~A.} \bibnamefont{Gershenfeld}}
  \bibnamefont{and} \bibinfo{author}{\bibfnamefont{I.~L.}
  \bibnamefont{Chuang}}, \bibinfo{journal}{Science}
  \textbf{\bibinfo{volume}{275}}, \bibinfo{pages}{350} (\bibinfo{year}{1997}).

\bibitem[{\citenamefont{Ladd et~al.}(2001)\citenamefont{Ladd, Goldman,
  Yamaguchi, Yamamoto, Abe, and Itoh}}]{ladd02}
\bibinfo{author}{\bibfnamefont{T.~D.} \bibnamefont{Ladd}},
  \bibinfo{author}{\bibfnamefont{J.~R.} \bibnamefont{Goldman}},
  \bibinfo{author}{\bibfnamefont{F.}~\bibnamefont{Yamaguchi}},
  \bibinfo{author}{\bibfnamefont{Y.}~\bibnamefont{Yamamoto}},
  \bibinfo{author}{\bibfnamefont{E.}~\bibnamefont{Abe}}, \bibnamefont{and}
  \bibinfo{author}{\bibfnamefont{K.~M.} \bibnamefont{Itoh}}
  (\bibinfo{year}{2001}), \eprint{quant-ph/0109039}.

\bibitem[{\citenamefont{Schulman and Vazirani}(1999)}]{schulman99}
\bibinfo{author}{\bibfnamefont{L.~J.} \bibnamefont{Schulman}} \bibnamefont{and}
  \bibinfo{author}{\bibfnamefont{U.~V.} \bibnamefont{Vazirani}},
  \bibinfo{journal}{Proc.\ 31st ACM\ Symp.\ on Theory of Computing} p.
  \bibinfo{pages}{322} (\bibinfo{year}{1999}).

\bibitem[{\citenamefont{Tycko}(1998)}]{tycko98}
\bibinfo{author}{\bibfnamefont{R.}~\bibnamefont{Tycko}},
  \bibinfo{journal}{Sol.\ State Nuc.\ Mag.\ Res.}
  \textbf{\bibinfo{volume}{11}}, \bibinfo{pages}{1} (\bibinfo{year}{1998}).

\bibitem[{\citenamefont{Lampel}(1968)}]{lampel68}
\bibinfo{author}{\bibfnamefont{G.}~\bibnamefont{Lampel}},
  \bibinfo{journal}{Phys.\ Rev.\ Lett.} \textbf{\bibinfo{volume}{20}},
  \bibinfo{pages}{491} (\bibinfo{year}{1968}).

\bibitem[{\citenamefont{Meier and Zakharchenya}(1984)}]{meier84}
\bibinfo{editor}{\bibfnamefont{F.}~\bibnamefont{Meier}} \bibnamefont{and}
  \bibinfo{editor}{\bibfnamefont{B.~P.} \bibnamefont{Zakharchenya}}, eds.,
  \emph{\bibinfo{title}{Optical Orientation}}, vol.~\bibinfo{volume}{8} of
  \emph{\bibinfo{series}{Modern Problems in Condensed Matter Science}}
  (\bibinfo{publisher}{North Holland}, \bibinfo{address}{Amsterdam},
  \bibinfo{year}{1984}).

\bibitem[{\citenamefont{Michal and Tycko}(1998)}]{michal98}
\bibinfo{author}{\bibfnamefont{C.~A.} \bibnamefont{Michal}} \bibnamefont{and}
  \bibinfo{author}{\bibfnamefont{R.}~\bibnamefont{Tycko}},
  \bibinfo{journal}{Phys.\ Rev.\ Lett.} \textbf{\bibinfo{volume}{81}},
  \bibinfo{pages}{3988} (\bibinfo{year}{1998}).

\bibitem[{\citenamefont{Michal and Tycko}(1999)}]{michal99}
\bibinfo{author}{\bibfnamefont{C.~A.} \bibnamefont{Michal}} \bibnamefont{and}
  \bibinfo{author}{\bibfnamefont{R.}~\bibnamefont{Tycko}},
  \bibinfo{journal}{Phys.\ Rev.\ B} \textbf{\bibinfo{volume}{60}},
  \bibinfo{pages}{8672} (\bibinfo{year}{1999}).

\bibitem[{\citenamefont{Farah et~al.}(1998)\citenamefont{Farah, Dyakonov,
  Scalbert, and Knap}}]{farah98}
\bibinfo{author}{\bibfnamefont{W.}~\bibnamefont{Farah}},
  \bibinfo{author}{\bibfnamefont{M.}~\bibnamefont{Dyakonov}},
  \bibinfo{author}{\bibfnamefont{D.}~\bibnamefont{Scalbert}}, \bibnamefont{and}
  \bibinfo{author}{\bibfnamefont{W.}~\bibnamefont{Knap}},
  \bibinfo{journal}{Phys.\ Rev.\ B} \textbf{\bibinfo{volume}{57}},
  \bibinfo{pages}{4713} (\bibinfo{year}{1998}).

\bibitem[{\citenamefont{Slichter}(1990)}]{slichter90}
\bibinfo{author}{\bibfnamefont{C.~P.} \bibnamefont{Slichter}},
  \emph{\bibinfo{title}{Principles of Magnetic Resonance}},
  vol.~\bibinfo{volume}{1} of \emph{\bibinfo{series}{Springer Series of Solid
  State Sciences}} (\bibinfo{publisher}{Springer Verlag},
  \bibinfo{address}{Berlin}, \bibinfo{year}{1990}).

\bibitem[{\citenamefont{Aonuma et~al.}(1993)\citenamefont{Aonuma, Sawa, Okano,
  Kato, and Kobayashi}}]{aonuma93a}
\bibinfo{author}{\bibfnamefont{S.}~\bibnamefont{Aonuma}},
  \bibinfo{author}{\bibfnamefont{H.}~\bibnamefont{Sawa}},
  \bibinfo{author}{\bibfnamefont{Y.}~\bibnamefont{Okano}},
  \bibinfo{author}{\bibfnamefont{R.}~\bibnamefont{Kato}}, \bibnamefont{and}
  \bibinfo{author}{\bibfnamefont{H.}~\bibnamefont{Kobayashi}},
  \bibinfo{journal}{Synth.\ Metal.} \textbf{\bibinfo{volume}{58}},
  \bibinfo{pages}{29} (\bibinfo{year}{1993}).

\bibitem[{\citenamefont{Schuerlein et~al.}(1995)\citenamefont{Schuerlein,
  Schmidt, Lee, Nebesny, and Armstrong}}]{schuerlein95}
\bibinfo{author}{\bibfnamefont{T.~J.} \bibnamefont{Schuerlein}},
  \bibinfo{author}{\bibfnamefont{A.}~\bibnamefont{Schmidt}},
  \bibinfo{author}{\bibfnamefont{P.~A.} \bibnamefont{Lee}},
  \bibinfo{author}{\bibfnamefont{K.~W.} \bibnamefont{Nebesny}},
  \bibnamefont{and} \bibinfo{author}{\bibfnamefont{N.~R.}
  \bibnamefont{Armstrong}}, \bibinfo{journal}{Jpn.\ J.\ Appl.\ Phys.\ 1}
  \textbf{\bibinfo{volume}{34}}, \bibinfo{pages}{3837} (\bibinfo{year}{1995}).

\end{thebibliography}

\end{document}